\newcommand{\eat}[1]{}
\newcommand{\stitle}[1]{\noindent{\bf #1}}
\documentclass[sigconf,nonacm]{acmart}
\usepackage{tcolorbox}
\usepackage{lipsum}
\usepackage{enumitem}
\usepackage[utf8]{inputenc}
\usepackage{mdframed}

\usepackage{listings}
\usepackage{colortbl}
\AtBeginDocument{%
  }

\setcopyright{acmlicensed}
\copyrightyear{2025}
\acmYear{2025}
\acmDOI{}
\acmConference[KDD Cup'25]{2025 KDD Cup Workshop}
\acmISBN{}

\begin{document}

\title{DB3 Team's Solution For Meta KDD Cup' 25}


\author{%
    \textsuperscript{1}Yikuan Xia\textsuperscript{*},
    \textsuperscript{1}Jiazun Chen\textsuperscript{*},
    \textsuperscript{1}Yirui Zhan
    \textsuperscript{1}Suifeng Zhao}
\author{%
    \textsuperscript{2}Weipeng Jiang,
    \textsuperscript{2}Chaorui Zhang,
    \textsuperscript{2}Wei Han,
    \textsuperscript{2}Bo Bai\textsuperscript{†},
    \textsuperscript{1}Jun Gao\textsuperscript{†}
}
\thanks {%
    \textsuperscript{*}These authors contributed equally to this research.\\
    \textsuperscript{†}Corresponding Authors
}

\affiliation{%
    \textsuperscript{1}
    \institution{Key Laboratory of High Confidence Software Technologies, CS, Peking University, China}
    \city{wfl00014@pku.edu.cn, \{chenjiazun, zhanyirui, suifengzhao\}@stu.pku.edu.cn, gaojun@pku.edu.cn}
    \country{China}
}

\affiliation{%
    \textsuperscript{2}
    \institution{Theory Lab, Central Research Institute, 2012 Labs, Huawei Technologies Co., Ltd, Shenzhen, China}
    \city{\{jiangweipeng, zhang.chaorui, harvey.hanwei, baibo8\}@huawei.com}
    \country{China}
}


\begin{abstract}
This paper presents the \textbf{db3} team's winning solution for the Meta CRAG-MM Challenge 2025 at KDD Cup'25. Addressing the challenge's unique multi-modal, multi-turn question answering benchmark (CRAG-MM), we developed a comprehensive framework that integrates tailored retrieval pipelines for different tasks with a unified LLM-tuning approach for hallucination control. Our solution features (1) domain-specific retrieval pipelines handling image-indexed knowledge graphs, web sources, and multi-turn conversations; and (2) advanced refusal training using SFT, DPO, and RL. The system achieved 2nd place in Task 1, 2nd place in Task 2, and 1st place in Task 3, securing the grand prize for excellence in ego-centric queries through superior handling of first-person perspective challenges.
\end{abstract}




\maketitle

\section{Introduction}

Recent advancements in artificial intelligence have highlighted the importance of multi-modal reasoning, especially in scenarios where systems must interpret and respond to both visual and textual information. The Meta CRAG-MM Challenge 2025~\cite{wang2025cragmmmultimodalmultiturncomprehensive} hosted on AIcrowd, aims to accelerate progress in this area by introducing a novel benchmark, called CRAG-MM, that encompasses both single-turn and multi-turn visual question answering tasks. Unlike traditional VQA datasets, CRAG-MM is designed to evaluate factual question answering in multi-modal contexts, integrating images with complex, multi-turn conversational flows.

Notably, \textbf{ego-centric images} are a major characteristic of this contest. In the context of the Meta CRAG-MM Challenge 2025, ego-centric images refer to images captured from a first-person perspective, typically using wearable cameras mounted on glasses. Unlike traditional third-person or static camera views, ego-centric images provide a direct visual record of the user's personal experience and interactions within their environment. This perspective enables the capture of rich contextual information, including hand-object interactions, dynamic activities, and real-time environmental changes. However, ego-centric images also present unique challenges, such as frequent occlusions, rapid viewpoint shifts, and complex backgrounds. Addressing these challenges is crucial for advancing machine perception and understanding of daily human activities from a truly immersive viewpoint, making ego-centric image analysis a key focus of this competition.

The Meta CRAG-MM Challenge 2025 is composed of the following three tasks:

\begin{enumerate}

    \item \textbf{Task 1:} Participants are required to answer single-turn questions using both image-indexed knowledge graphs as context.

    \item \textbf{Task 2:} This task also involves single-turn questions but additionally introduces a web search knowledge source.

    \item \textbf{Task 3:} A multi-turn retrieval-augmented generation (RAG) task, where systems must engage in dialogue-like interactions, retrieving and integrating information from multiple sources to generate coherent responses.

\end{enumerate}

The author's team, db3, participates in the contest and achieves second place, second place, and first place in the three tasks, respectively, resulting in winning the grand prize in ego image total score. This paper describes the author's solution to the three tasks. 

The two main challenges of this contest are 1. How to deal with the retrieval information, both the multimodal part and the multi-turn conversation part. 2. How to control hallucinations. As the metrics of the contest involve punishing answering incorrect answers, we have to train the model to output "I don't know" on queries hard for them to answer. Therefore, we describe our solutions in two parts: the retrieval part and the hallucination control part. In this paper, in addition to presenting our final submitted solution, we also describe the various approaches and experiments we attempted during the development process. Our code is available on GitLab~\footnote{https://gitlab.aicrowd.com/jiazunchen/db3-team-s-solution-for-meta-kdd-cup-25}.

In the remainder of the paper, we first introduce some basic constructions we make for this contest \label{sec:basic}. We discuss the retrieval part in Sec.~\ref{sec:1}, and we discuss the hallucination control part in Sec.~\ref{sec:23}. We discuss the checkpoint selection and ensemble tricks we use in this contest in Sec.~\ref{sec:trick}. We conclude our work and look into future works in Sec.~\ref{sec:conclusion}.

\begin{figure*}
    \centering
    \includegraphics[width=1\linewidth]{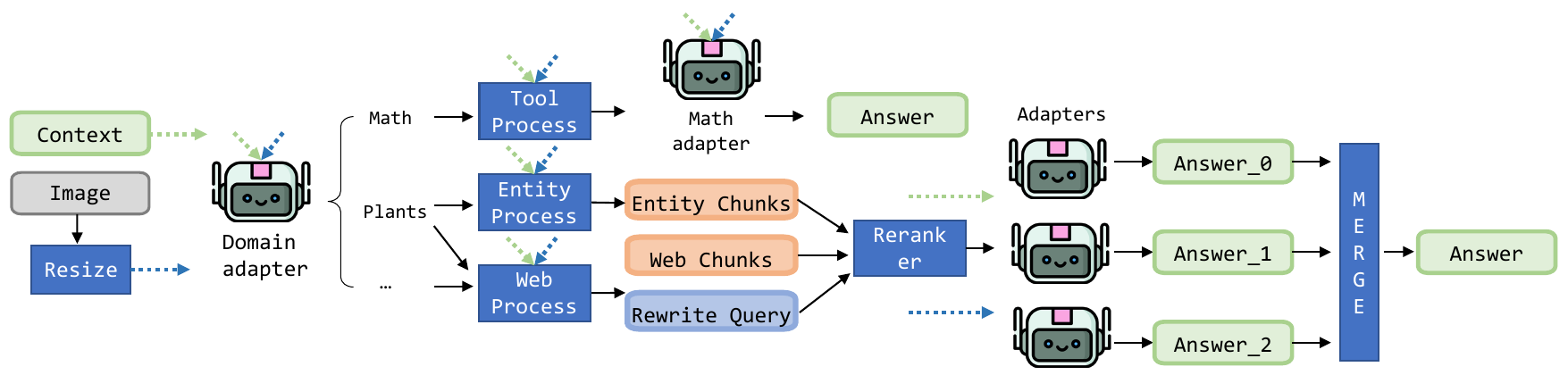}
    \caption{Pipeline for Different Tasks: A domain adapter is first applied to classify the query domain for different pipelines, e.g., a math tool for math problems, a retrieval through entity chunks module for plants, and a rewriting and retrieving from web chunks module for all other domains. After a reranker module for all retrieved text chunks, we obtain multiple answers using trained hallucination control adapters, and these answers are ensembled to obtain a final output answer.}
    \label{fig:enter-label}
\end{figure*}

\section{Basic Constructions.}
\label{sec:basic}
We construct some basic modules for this contest that are useful for all the following procedures.

\stitle{LLM judge.} Since the online judge and prompt for this contest are not available, we construct a local LLM judge for this contest. We observe that using different prompts and LLM judges can have a huge difference in the results, and some of our training process relies on the LLM judge's results. We enumerate some combinations of the prompts and LLM APIs we use locally and reconcile the local scores with the online scores we get. The following prompt applied with GPT-4o-mini achieves the relatively most similar results with online evaluation results. The prompt used is as follows:

\begin{lstlisting}[label={lst:structured_data_processing}]
You are an expert evaluator for question answering systems. 
Your task is to determine if a prediction correctly answers a question based on the ground truth.
    
Rules:
1. The prediction is correct if it captures all the key information from the ground truth.
2. The prediction is correct even if phrased differently as long as the meaning is the same.
3. The prediction is incorrect if it contains incorrect information or is missing essential details.


Question: {query}
Ground truth: {ground_truth}
Prediction: {prediction}

Output only 'true' or 'false' to indicate if the prediction is correct.
\end{lstlisting}

Noting that this prompt is not the best one that follows human evaluation. Through our observation, this prompt and online evaluation is much stricter than human evaluation, resulting in many of the results that should be correct being assigned as wrong during online evaluation. However, we have to choose this prompt to pursue a better score during the online evaluation.

\stitle{Train/Validation Split.} To perform checkpoint selection and other validation-based techniques, we split the original training data into distinct training and validation subsets. Given the diverse query types and domains present in the dataset, it is crucial to ensure that both splits are representative of the overall data distribution. To achieve this, we employ a stratified splitting strategy based on question types and domains, ensuring that each category appears proportionally in both the training and validation sets. Specifically, we allocate about 3,000 of the samples to the training set and 800 to the validation set, maintaining an 8:2 ratio across all types and domains. This approach mitigates the risk of overfitting to specific data characteristics and supports more reliable model evaluation and checkpoint selection.

\begin{figure}
    \centering
    \includegraphics[width=0.9\linewidth]{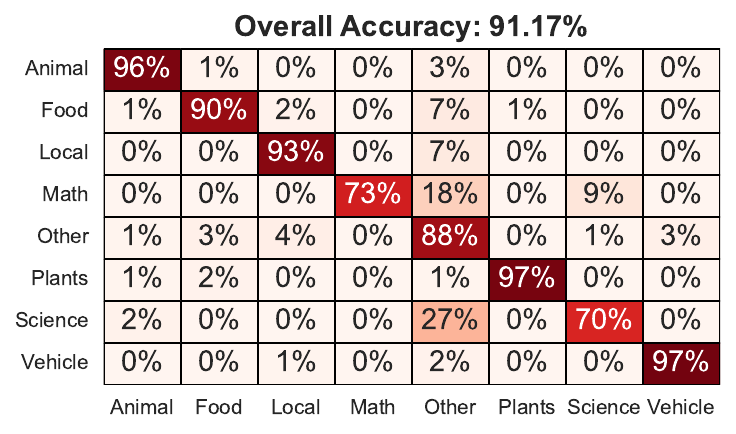}
    \caption{Confusion Matrix of Domain Prediction}
    \label{fig:conf}
\end{figure}

\stitle{Domain Prediction.} We rely on the domain splits to apply different pipelines for different domains. Since some of the domains have blurred semantics, we combine some of the originally split domains and have these final domains for classification: vehicle, plant, local, math, science, food, animal, and other. We re-annotate the queries using the new classes, and we tune Llama 3.2-VL for domain classification. The performance of this module is acceptable, as the classification accuracy of this model is 91.17\% on the validation set, and many of the wrongly assigned queries actually can fit in many classes. The confusion matrix is presented in Fig.~\ref{fig:conf}.

\section{The Retrieval Component of the Solution}
\label{sec:1}

In this section, we will propose the retrieval component of our solution. In this contest, the query is always associated with a query image. Therefore, we need to deal with both the query image and the text query. Specifically, we will present pipelines involving retrieving from the image-indexed knowledge graph in Task \#1, the text-indexed web source in Task \#2, and the multi-turn context in Task \#3. Additionally, two techniques that are useful for all three tasks are presented in this section: the OCR extraction and the tool solution for math problems.

\subsection{Retrieval Pipeline for Image-indexed Knowledge Graph.}

In Task 1, an image-indexed knowledge graph is provided as the information source. Specifically, an image is linked with a CLIP~\cite{radford2021learning} index and is associated with a structured segment of text, which serves as a description of the entity represented by the image. The information useful for answering the question is mostly based on the information in the structured text; therefore, the task involves identifying the correct entity the query is about in the knowledge graph.

\stitle{Retrieving and Reranking through Image.} The most direct solution is to retrieve and rerank through the query image and the index images. The most related images are first retrieved through the CLIP index, and the related text information is used as the retrieved information. The key challenge here is the quality of the retrieval results. Through experiment, we observe that directly applying the top 1/3/5 results of this source through the image index harms the final QA accuracy. We applied the following techniques to improve the retrieval quality.

\textbf{Grounding In the Query Image.} Ego-centric images consist of a large portion of the query images. The distribution of ego-centric images differs a lot from the index images. The ego-centric images are mostly taken in everyday life and contain much irrelevant background, while the index image is often a Wikipedia-style image concentrating on the entity. Therefore, the CLIP embeddings of the query image and correct index image are very likely to differ a lot. For example, if the query is about a car parked in front of a building, the CLIP embedding of the query image will be more similar to a building's CLIP embedding. We have to clip the car part of the query image to retrieve the correct car entity in the knowledge graph. Therefore, we introduce grounding in the query image to resolve this problem. We leverage the open-source Grounding-DINO~\cite{liu2024grounding} model to deal with this task. The Grounding-DINO model takes in a short text and an image as input. The short text serves as a prompt describing the object or region of interest (for example, “the plant in the middle” or “a black car”). The model then processes both the image and the text to identify and localize the region in the image that best matches the given description.

Through our annotation, we find that with the appropriate text input, the Grounding-DINO model can
localize the correct entity in most cases, and through statistics, the correct entities appear more frequently in the retrieved top entities. However, the text input is quite hard to obtain. The ideal input is related to the query and the query image, and it's hard to prompt the VLM to output the ideal describing text having the same quality as the human-annotated ones. In practice, we prompt VLMs to generate the describing text. The Dino prompt we used is as follows:

\begin{lstlisting}[label={lst:structured_data_processing}]
Image: {}
Given this image, a query, your task is to simply describe the object in the image.    
Query: {}
Output only simple object names in phrases, do not output a sentence.
Do not answer the query, just output the object name appearing in the image, not the answer or answer entity.
\end{lstlisting}

Alternatively, we can use a certain phrase for each domain, e.g., Car, Plant, Building,..., as the DINO phrase to minimize the LLM inference cost.

\textbf{Reranking through Image.} In the contest retrieval scenario, retrieving through CLIP embeddings is highly inaccurate. Following human practice, checking whether two images refer to the same entity is essential. Therefore, we propose reranking by comparing the query and index images. We use the following prompt to let VLMs judge whether the two images belong to the same entity. We use the image rerank prompt to do this task:
\begin{lstlisting}[label={lst:structured_data_processing}]
Image: {query image},{index image}
Given two images, the first one is a query image, the second one is an image about an entity, a query about the first image, descriptions about the second image, your task is to determine whether the query about the first image is about the entity in the second image.    
Query: {}
Description: {}
If the entity in the second image appears in the first image, output Yes, otherwise, output No.
\end{lstlisting}
We only preserve the related items (with judge output yes) in the top index. Through experiment, we observe that using this prompt with a powerful VLM, e.g., GPT-4o, can achieve considerable boosts in performance. Since only a relatively small VLM can be used in the contest, we aim to distill Llama 3.2-VL to achieve this ability. 

We sample the top 5 candidate items for each query in the training set using our image-based grounding and retrieval pipeline. Each candidate is verified by a powerful VLM (e.g., GPT-4o) to determine whether the entity in the index image matches the entity grounded in the query image. A candidate is labeled as relevant only if this condition is satisfied. This process resulted in a distillation dataset of approximately 15,000 candidate-query pairs, within which roughly 500 positive samples were identified for Task 1. The base VLM is subsequently trained on these data to classify the relevance of candidate items. To address class imbalance, we employ negative sampling during training to construct mini-batches with an approximately 1:1 ratio of positive to negative samples. This balanced sampling strategy ensures that the model receives representative supervisory signals from both relevant and irrelevant instances at each optimization step, thereby mitigating bias toward the majority class.

In our experiment, we observe that reranking with a powerful VLM can largely boost the final QA results (by 10\%), and using a distilled Llama 3.2-VL can boost the QA performance by 2\%. Since the boost is tiny, and we try this technique in the early stage of the contest, we don't have time to add it to the final pipeline. 
\begin{figure}
    \centering
    \includegraphics[width=1\linewidth]{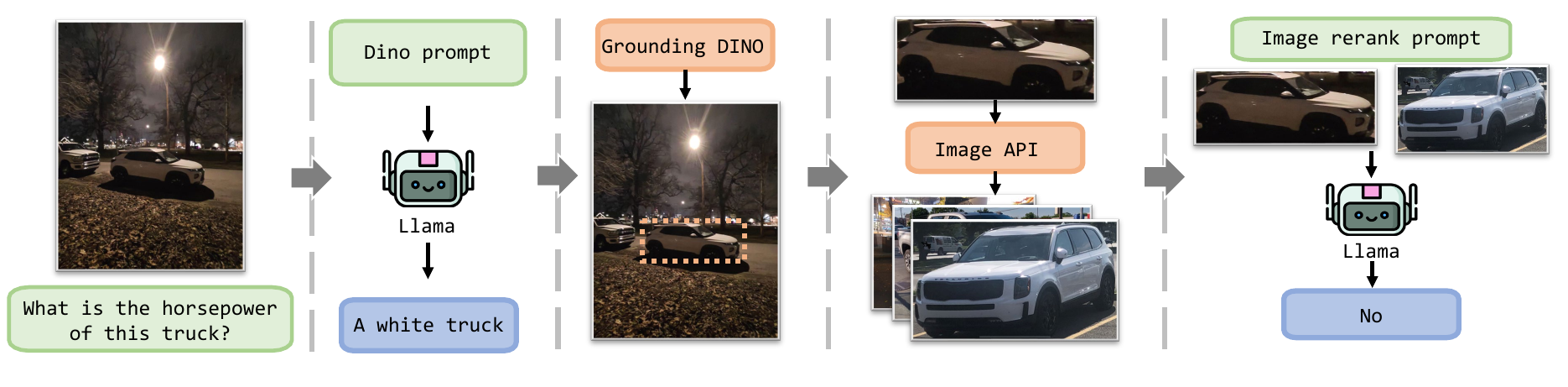}
    \caption{Grounding in the Query Image + Rerank through Image}
    \label{fig:enter-label}
\end{figure}

\stitle{Retrieving and Reranking through Text.} Since we observe severe difficulty in utilizing the top-k candidates through CLIP image embedding, we propose retrieving and reranking through text. In many cases, the VLM itself can identify the correct entity name the query is about. We can retrieve the relevant attributes about this entity from the image-indexed knowledge graph. We made two attempts at retrieving through text: the entity name approach and the merged text query approach.
 
\textbf{The entity name approach.} The entity name approach means first extracting the desired entity name from the query and query image, and then identifying the same entity from the knowledge graph. For example, the query asks about when this plant blooms; the entity name will be the plant's botanical name, which is expected to exist in the knowledge graph.

In the first step, we prompt the VLM to extract the query entity's name. We use the entity prompt to conduct this task.

\begin{lstlisting}
Image:{}
Given an image and a query about it, your task is to extract the entity's name the query is about.
If the entity is an animal, output its  scientific name.
If the entity is a plant, output its full botanical name.
If the entity is food, output the specific dish or ingredient name.
If the entity is a location or landmark, output its common name.
If the entity is a mathematical concept or symbol, output its formal name.
If the enityt is about a vehicle, you should output the vehicle's brand and model.
For other entities, output the most specific and commonly recognized name.
Query: {}
You should output the entity's name directly.
\end{lstlisting}

The prompt consists of detailed entity name instructions that match the entities' names in the knowledge graph. Similarly, prompting strong VLMs, e.g., GPT-4o, can extract more correct entity names compared to prompting Llama 3.2-VL. On one hand, GPT-4o possesses more internal knowledge. On the other hand, GPT-4o can follow the instructions better. We also tune Llama 3.2-VL using the ground truth label name to learn the format and follow the instructions better. (The ground truth label is obtained by similar prompts with the ground truth answer. As the ground truth label usually appears in the ground truth answer, most of the GPT-4o extraction results are identical to the ground truth entity name.)

In the second step, we use the retrieving-through-image method to extract a large number of candidates (e.g., 1000), and the generated entity name is compared with the entity names of the candidates. The entity names are tokenized and stemmed, and only if the processed token set is identical or contains the other are the two entity names considered related.

Through experiment, we find that using GPT-4o, 40\% of the entity name can be extracted correctly. The ratio is 20\% for Llama 3.2-VL. The potential of the QA system can be boosted by 10\% using the GPT-4o result, and the result is 5\% for the Llama 3.2-VL results. Unfortunately, we didn't have enough time to train a model that performed well in all domains. As a result, we could only apply this solution in the field of plants.

Notably, even powerful VLMs like GPT-4o can only identify 40\% of the entity's desired names. This indicates that for the hard cases like vehicles, plants, and food, state-of-the-art VLMs still can't recall their names, as different models of cars or different plants in the same class look quite similar. During the contest, we try to search for open-source models to identify a car's brand and model or plants' botanical names. However, it turns out that there are no such open-source models, and a system that can deal with these needs requires a highly sophisticated database building and model training.

\begin{figure}
    \centering
    \includegraphics[width=1\linewidth]{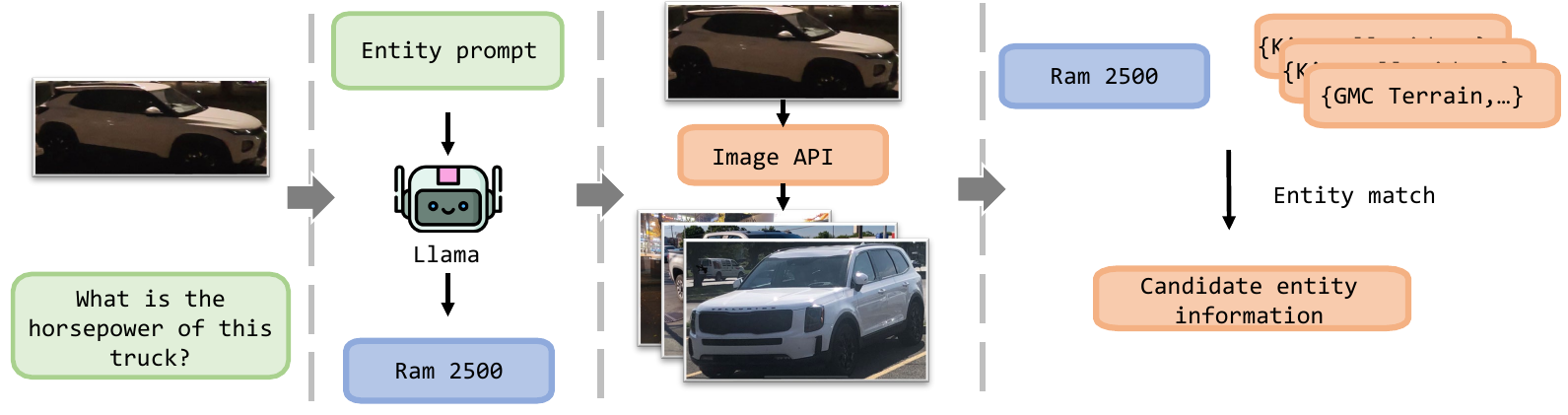}
    \caption{Retrieving and Reranking through Text.}
    \label{fig:enter-label}
\end{figure}

\textbf{Query using merged text query.} A merged text query is a pure text query that contains all the query information in the original text query and the query image. We will propose the full merged text query rewrite process in Sec. 3.2 in detail. Suppose we have the merged query here. We follow the same steps of retrieving a large number of candidates (e.g., 1000), and we convert the structured text information of the candidates to a text embedding retrieval base. 

Specifically, the attribute-value pair in the entity's structured text is converted into separate text outputs. For example, the Volkswagen Beetle entity has an attribute, end of production year: 2019. This attribute value pair will be converted to the end of the production year of Volkswagen Beetles, which is 2019. These sentences, each representing one attribute of an entity, are further selected using rerank models, e.g., BGE-reranker-v2-m3.

Through experiment, we find that the potential of the QA system can be boosted by 20\% using this technique. We successfully integrated this technique into our hallucination control module, and the overall performance is slightly boosted for the plant category.

\subsection{Retrieval Pipeline for Text-indexed Web Source.}

In Task 2, a text-indexed web page source is provided as an information source. Specifically, the chunked web page content is linked with a text embedding index. The difficulty in this task lies in how to rewrite the query into a merged text query that has the same modality as the web page text, which can be retrieved through the text embedding. For example, for the query, When does this car stop production?, and the query image containing a Volkswagen Beetle, we have to rewrite the original query to When does the Volkswagen Beetle stop production? We have to substitute the pronouns in the original query with information from the query image and rewrite a merged query that can be answered individually without the query image.

\begin{figure}
    \centering
    \includegraphics[width=1\linewidth]{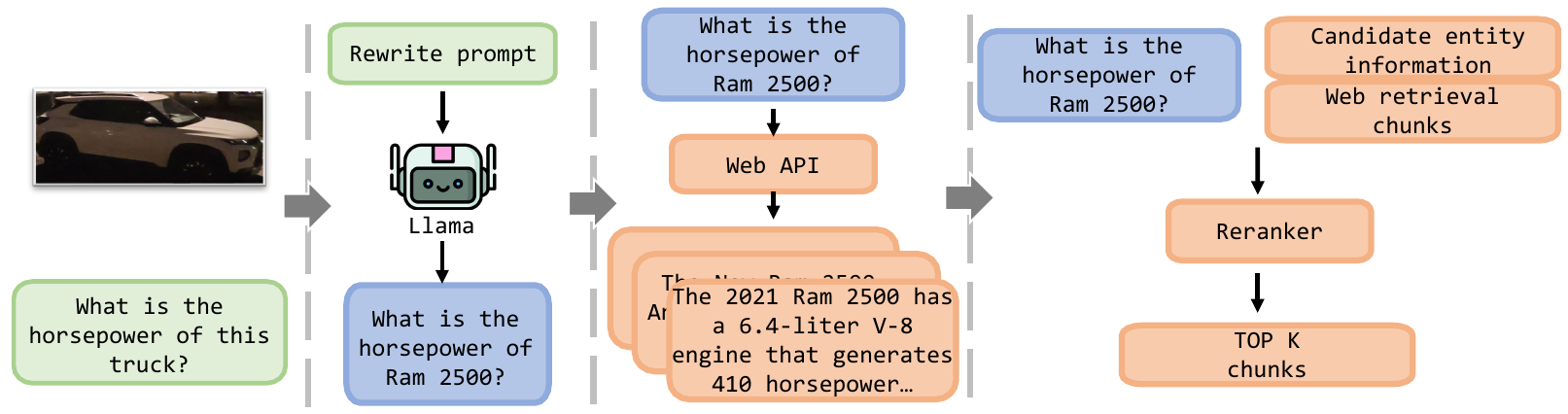}
    \caption{Retrieval Pipeline for Text-indexed Web Source.}
    \label{fig:enter-label}
\end{figure}

\stitle{SFT tuning for Merge Query Rewrite.} We can prompt VLMs to generate a merged query. We use the rewrite prompt to rewrite a merged query:

\newpage

\begin{lstlisting}[label={lst:structured_data_processing}]
Image:{}
###Task 
You are an expert at converting visual questions into effective search queries. 
Your goal is to create a comprehensive search query that will help find the most relevant information. 
For each image-based question, you must create a search query that combines: 
1. Key visual elements from the image (objects, text, logos, scenes, actions, etc.) 
2. The core question being asked 
3. Potential answer terms or relevant context 
For example: 
- If asking about a logo: include company name, industry, and visual description 
- If asking about an object: include its appearance, category, and possible brands/models 
- If asking about an event/scene: include location hints, activities, and time period clues
Query: {}
Answer: {} //When cheating, it will provide the answer.
\end{lstlisting}

We first prompt the powerful GPT-4o to generate the merged query rewrite. As we have mentioned before, even GPT-4o can only identify half of the entities in this contest. Moreover, the instructions we make in the prompt are still not 100\% followed by GPT-4o. This is worse while using Llama 3.2-VL for rewriting. Here we list the potentially correct answering case in Task 2 using the rewrite system. As we can see, the GPT-4o rewrite system can boost answer potential by \~13\%, while the Llama 3.2-VL rewrite can only boost it by 7\%.

\begin{table}[h]
\centering
\caption{Comparison of Different Rewriting Methods}
\label{tab:rewrite}
\begin{tabular}{lc}
\toprule
Method & Score (\%) \\
\midrule
Ori & $\sim$33 \\
Llama Rewrite & $\sim$40 \\
GPT-4o Rewrite & $\sim$46 \\
Llama-distill Rewrite & $\sim$44 \\
GPT-4o Cheat Rewrite & $\sim$60 \\
Llama-distill-cheat Rewrite & $\sim$44 \\
\bottomrule
\end{tabular}
\end{table}

To solve the difficulty of entity extraction, we introduce a cheated version of rewrite. In the prompt we listed above, we additionally add the ground truth answer. As the ground truth entity mostly appears in the ground truth answer, providing the ground truth answer can guide the VLM to include the ground truth entity in the rewrite query. As we can see, the cheated version achieves 60\% of the potential correct score.

We tune the Llama 3.2-VL base model to enhance its query rewriting capability. For the training data, we utilize both cheated and non-cheated rewrite queries, but critically, we only employ those queries for which GPT-4 generated a correctresponse as the ground-truth label to supervise the fine-tuning process of Llama 3.2-VL. This ensures that the model learns from accurate and reliable demonstrations, improving its ability to generate effective query rewrites. The results are also presented in Table 1. As we can see, regardless of using the cheated or non-cheated version, the boost of potential performance is \~4\%. The performance is close to the GPT-4o rewrite performance (~3\% down); therefore, the result is rather satisfying. We choose one of the checkpoints in the Llama-distill checkpoints as the final rewrite module we use in the submitted version.

\stitle{RL Tuning for Merge Query Rewrite.} During the contest, we also try using RL to tune this rewrite module. We try two different approaches, the DPO approach and the RL (GRPO) approach.

\textbf {DPO approach.~\cite{rafailov2023direct}} We follow the normal procedure of DPO training. We construct the training data as follows. Following the best SFT checkpoint we obtain in the last subsection, we sample 5 rewrite merge queries using high temperature. The better pairs in DPO are the rewrite queries that can produce the context, making the QA result correct, and the worse pairs in DPO are the rewrite queries that produce the context, making the QA result incorrect.

\textbf{RL (GRPO) approach.~\cite{shao2024deepseekmath}} We use the GRPO algorithm in RL training for query rewrite. Similarly, the merge query is rewarded if the context helps answer correctly. Conversely, the merge query is punished if the context makes the answer wrong.

\textbf{Outcome.} As we have mentioned, due to the lack of sufficient internal knowledge, the space for further improvement through RL is limited. The gap between tuned Llama and GPT-4o is only 3\%. In our experiment, we fail to obtain a better rewrite checkpoint than the original SFT-tuned checkpoint. Therefore, our submitted version is based on SFT-tuning.

\stitle{Retrieval through Text.} For simplicity, we directly use the preprocessed web content instead of the original HTML content. After obtaining the rewritten merged query, we retrieve the text chunks using the bge-large-en-v1.5 index. Conventionally, we further use BGE-reranker-v2-m3 to sort the candidate chunks more accurately~\cite{chen2024bge}.

\subsection{Retrieval Pipeline for Multi-turn Conversation.}

In Task 3, the information source is the same as the source in Task 2. The difference is the introduction of a multi-turn conversation. In this subsection, we present our retrieval pipeline for multi-turn conversation.

\stitle{One-step Context.} Though some queries may require longer context, we observe in our experiment that, provided with the ground-truth QA, the QA performance of using one-step context is almost the same as the result of using full context. Therefore, for the simplicity of sampling and training, we only use the last-step context in the contest.

\stitle{Merge Query Rewrite with Context.} Since many of the queries require context QAs to specify the entity the query is about, we add the query QAs into the merge query rewrite prompt. The prompt we use is as follows:

\newpage

\begin{figure}
    \centering
    \includegraphics[width=1\linewidth]{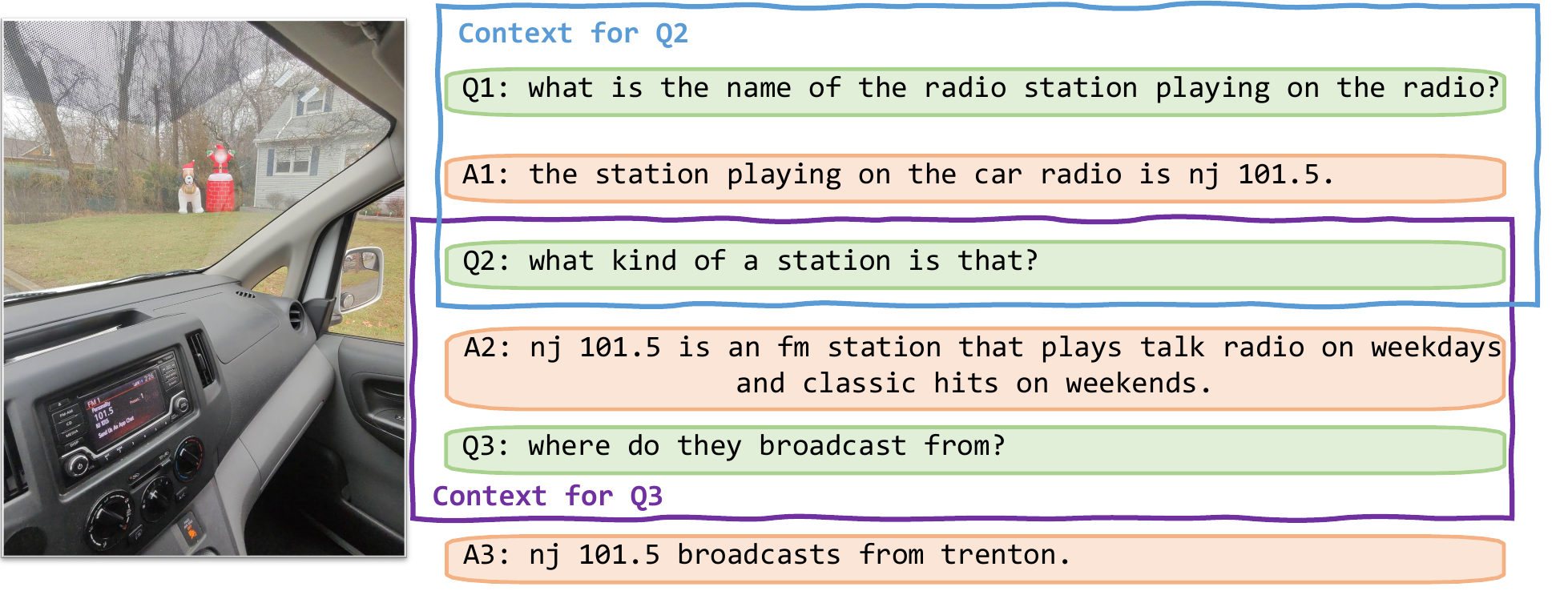}
    \caption{One-step Context for Multi-Round QA.}
    \label{fig:enter-label}
\end{figure}
\begin{lstlisting}
Image:{}
###Task 
You are an expert at converting visual questions into effective search queries. 
The current query is a part of multi-turn conversation. You should use the history conversation to make sure what the current query is about.
Your goal is to create a comprehensive search query that will help find the most relevant information for the currecnt query. 
For each image-based question, you must create a search query that combines: 
1. Key visual elements from the image (objects, text, logos, scenes, actions, etc.) 
2. The core current question being asked 
3. Potential answer terms or relevant context 
For example: 
- If asking about a logo: include company name, industry, and visual description 
- If asking about an object: include its appearance, category, and possible brands/models 
- If asking about an event/scene: include location hints, activities, and time period clues 
Query: {}
Context: {}

\end{lstlisting}

The tuning process is the same as the process without a multi-turn conversation.

\stitle{Sampling Strategy.} To achieve the best final score, a hallucination control component has to be employed. However, in merge query rewrite training, we need to deploy the control component, as sufficient QA information has to be included in context.

\subsection{OCR from the Query Image.} Through observation, we find that many of the queries are directly about the text in the image. Therefore, we suspect employing a separate OCR module can boost the performance for queries of this kind. Through experiment, we find that many more queries can be answered correctly using OCR results from GPT-4o.

However, we find that using VLM models to extract text has a severe drawback. A large number of queries involve a large amount of text from a book. Conducting OCR using VLMs requires generating many tokens, making this approach unavailable in this contest.

We also try using OCR tools like PaddleOCR~\cite{cui2025paddleocr30technicalreport}. However, we find that the extraction results are not as satisfying as using VLMs, particularly in the ego-centric images. Considering all the difficulties, we do not include an OCR component in our pipeline. However, we believe this component can be useful, as the base VLM normally can't focus on too much text in the image.

\subsection{Solve Math Problems Using Tools.} Math problems are always tricky for LLMs/VLMs, and using tools is a standard solution for this. Therefore, we employ using tools to solve the math problems. Through observation of the contest data, we conclude there are three types of major problems in math: calculation and simplification of numbers and variables, base conversion, and chemical formula balancing. We construct tools for these three types of queries, and we provide APIs for these tools. We prompt the VLM to follow the provided tool API. To solve the instruction-following problem, we construct \~50 tool examples and tune the base VLM. The final version of our math module can solve most of the math problems correctly in the training data of the contest.

\begin{figure}
    \centering
    \includegraphics[width=1\linewidth]{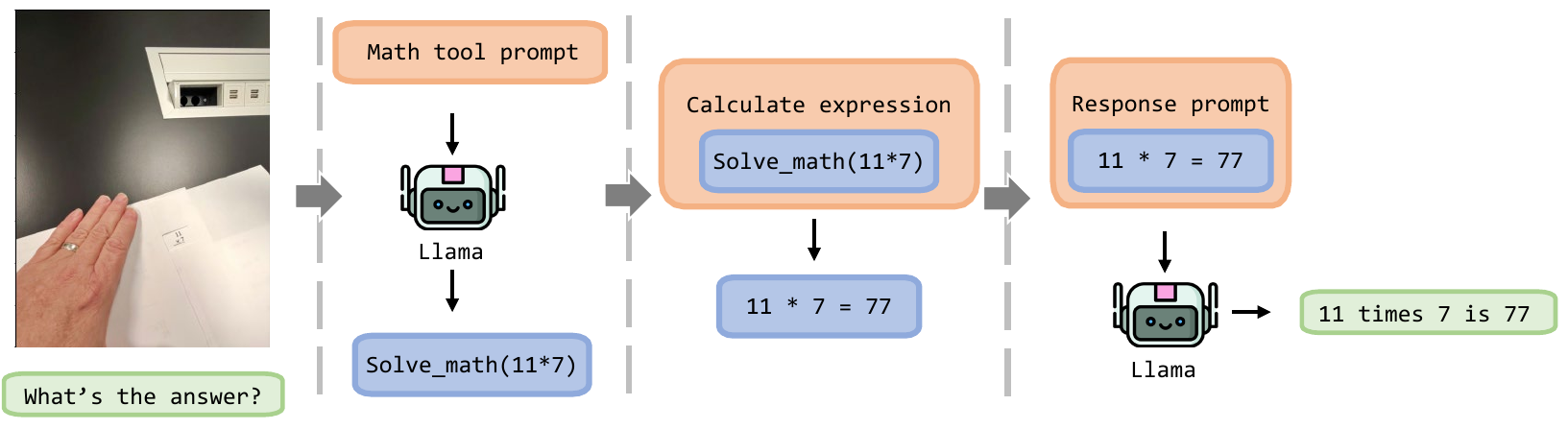}
    \caption{Solve Math Problems Using Tools.}
    \label{fig:enter-label}
\end{figure}

\section{Hallucination Control Component of the Solution}
\label{sec:23}

LLMs/VLMs inevitably encounter queries that fall outside their reliable knowledge scope. During the contest, we are awarded 1 point for answering correctly and penalized 1 point for answering incorrectly; therefore, we have to control hallucinations to achieve higher scores.

The hallucination control component therefore, has two equally important goals:

\begin{enumerate}

    \item \textbf{Maximize the proportion of correct answers.}

    \item \textbf{Minimize the proportion of wrong answers by refusing when necessary.}

\end{enumerate}

\subsection{Answerability Estimation}
\label{sec:answerability}

Whether a query is answerable by the current retrieval and answering system can be determined by the correctness of its generated answers. We denote the queries that are hard for the current system to produce a correct answer as unanswerable.

\subsection{Refusal Training Pipeline}
\label{sec:refusal-training}

\stitle{Supervised Fine-Tuning (SFT)}

\begin{itemize}
    \item \textbf{Unanswerable queries}: label $\rightarrow$ \texttt{I don't know}.
    \item \textbf{Answerable queries}:  ground-truth reference answers 

\end{itemize}

\stitle{Direct Preference Optimisation (DPO)}
We build pairwise preferences $(\text{better},\text{worse})$:
\begin{itemize}
    \item \textbf{Answerable queries}:  
          better $=$ correct answer;  
          worse $=$ \texttt{I don't know}.
    \item \textbf{Unanswerable queries}:  
          better $=$ \texttt{I don't know};  
          worse $=$ hallucinated answer.
\end{itemize}

\stitle{Reinforcement Learning (GRPO)}
We adopt GRPO because it eliminates the value-function critic, reducing instability, yet the reward definition applies equally to PPO:
\[
r =
\begin{cases}
+k, & \text{correct answer};\\
0,  & \text{\texttt{I don't know}};\\
-1, & \text{incorrect answer}.
\end{cases}
\]
With $k{=}1$, the expected return is identical to the Refusal Score.

We conduct comprehensive experiments with all three approaches— SFT, DPO, and GRPO. The results indicate that the SFT approach performs poorly on this task, achieving a negative score. For Task 1, both DPO and GRPO yield comparable final scores of approximately 5\%. However, we observe notable behavioral differences: models trained with DPO tend to be more conservative, exhibiting a high missing rate (70\%-90\%) as they frequently respond with "I don‘t know.". In contrast, models trained with GRPO adopt a more proactive response strategy, demonstrating a lower missing rate (60\%-80\%) and showing a greater willingness to attempt answers. Additionally, each method displays distinct strengths across different domains. Given these complementary characteristics, we subsequently explore an ensemble strategy to leverage the advantages of both methods.

\section{Checkpoint Ensemble Trick}
\label{sec:trick}

Since the hallucination control process is tricky and highly unstable, selecting and combining the best checkpoints became quite important. In this section, we conclude the tricks we employ in this contest.

\stitle{Checkpoint Candidate Pool.} We collect the checkpoint results in many training trials under different settings to form a checkpoint candidate pool. The checkpoints with poorer results are neglected in the pool. All the results on the validation set are recorded so that further processes in the candidate pool do not require re-evaluation.

\stitle{Checkpoint Ensemble and Selection.} There are many ways of ensembling checkpoints. We list some of them:

\textbf{Ensemble according to Domain.} Since the difficulty of different domains varies a lot, we can control the QA checkpoints or even block answering according to the domain information. The strategy here can be selecting the best checkpoints on each domain, and if no checkpoint can achieve positive scores on this domain, we block the answers on this domain directly.

\textbf{Ensemble according to Equivalence.} We conduct equivalence clustering on all the answers of different checkpoints, and we select the answers supported by the most checkpoints. We can enumerate the checkpoints in the candidate pool to get the optimal checkpoint subset.

\textbf{Mixed Ensemble of Domain and Equivalence.} We can apply a mixed strategy, which means enumerating checkpoint combinations on each domain and selecting different optimal checkpoint combinations for each domain.

\stitle{Pros and Cons.} Pros: The ensemble process helps us boost plenty of performance on the final score.

Cons: Since many checkpoints are involved, it requires multiple rounds of inference during the test, making it hard to control the time limit. Unfortunately, many strategies that work well locally fail to work online, even if we manage to control the time limit carefully. This potentially harms the final score of the solution. 

Another major drawback is that conducting such a large-scale selection in the checkpoints creates a large gap between the local evaluation results and the online evaluation results. Our enumeration results show huge improvements locally, while it turns out to be heavily overfitting. Our final version also differs from the complex combinations for each domain, because the less selection is made, the less overfitting is observed.

\section{Conclusion}
This paper presented \textbf{db3}'s comprehensive solution for the Meta CRAG-MM Challenge 2025, which secured top rankings across all tasks and won the grand prize for ego-centric queries. Our key innovations include:

1. \textbf{Domain-Adaptive Retrieval}: We developed specialized pipelines for different data modalities, most importantly, merged query rewriting for text-based retrieval, significantly improving context relevance.

2. \textbf{Hallucination Control}: Through multi-stage training (SFT, DPO, RL) with refusal optimization, we created models that reliably output "I don't know" for unanswerable queries while maximizing correct responses.

Despite our success, limitations remain in fine-grained entity recognition and OCR integration for text-heavy images. Future work should explore dedicated recognition models and optimized OCR-VLM pipelines. Our solution demonstrates that combining task-specific retrieval with rigorous hallucination control is essential for reliable multi-modal QA systems, particularly for challenging ego-centric scenarios. 
\label{sec:conclusion}

\section{Acknowledgement}

We extend our sincere gratitude to the experts from Huawei's Theory Lab at the Central Research Institute, 2012 Labs, for their invaluable technical discussions and suggestions throughout this project.
This project is funded by NSFC (No. 62272008).

\bibliographystyle{ACM-Reference-Format}
\bibliography{sample-base}

\appendix
\begin{table*}[ht]
\caption{LoRA and FineTuning Hyperparameters.}
\resizebox{0.8\textwidth}{!}{
\begin{tabular}{l|l}
\hline
Name                             & Value                                                                                                                    \\ \hline
LoRA\_alpha                     & 16                                                                                                                       \\
LoRA\_dropout                   & 0.1                                                                                                                      \\
LoRA\_r                         & 8                                                                                                                        \\
target\_modules                 & \begin{tabular}[c]{@{}l@{}}{[}"k\_proj", "q\_proj", "v\_proj",  "up\_proj", "down\_proj", "gate\_proj"{]}\end{tabular} \\
bias                            & "none"                                                                                                            
\\
freeze\_vit  & True
\\

max\_seq\_length                & 2048 for task1 / 4096 for other tasks                                                                                                                 \\
per\_device\_train\_batch\_size & 4                                                                                                                        \\
gradient\_accumulation\_steps   & 4                                                                                                                        \\
optim                           & "adamw\_hf"                                                                                                              \\
learning\_rate                  & 2e-4                                                                                                                     \\
max\_grad\_norm                 & 0.3                                                                                                                      \\
scheduler                       & "cosine", warm\_up\_ratio = 0.1                   \\        
epoch   &3
\\
num\_generations (GRPO) & 8
\\ beta (KL coefficient) & 0.1 (DPO) / 0.04 (GRPO)
\\ 

\hline

\hline
\end{tabular}
} 
\label{tab:my-table}
\end{table*}

\section{LoRA and FineTuning Hyperparameters.}

The LoRA and Finetuning hyperparameters used in tuning the base model and API generation module is listed in Tab.~\ref{tab:my-table}:

\end{document}